\documentclass[sn-mathphys]{sn-jnl}

\jyear{2023}%

\usepackage{bm}

\begin{document}
\title[Verification of Wiedemann-Franz law in silver with moderate RRR]
{Verification of Wiedemann-Franz law in silver with moderate residual resistivity ratio}
\author[1,2]{\fnm{Marijn} \sur{Lucas}}
\author*[1]{\fnm{Lev V.} \sur{Levitin}}\email{l.v.levitin@rhul.ac.uk}
\author[1]{\fnm{Petra} \sur{Knappov\'{a}}}
\author[1]{\linebreak\fnm{J\'{a}n} \sur{Ny\'{e}ki}}
\author[1]{\fnm{Andrew} \sur{Casey}}
\author[1]{\fnm{John} \sur{Saunders}}
\affil[1]{\orgdiv{Department of Physics}, \orgname{Royal Holloway University of London}, \orgaddress{\street{Egham Hill}, \city{Egham}, \postcode{TW20 0EX}, \state{Surrey}, \country{United Kingdom}}}
\affil[2]{now at \orgname{Microsoft Quantum Lab Delft}, \orgaddress{\street{Lorentzweg} 1, University~of~Delft, \postcode{2628CJ}, \city{Delft},
\country{the Netherlands}}}

\date{25 July 2023}


\abstract{Electrical and thermal transport were studied in a vacuum-annealed polycrystalline silver wire with residual resistivity ratio 200\,-\,400,
in the temperature range 0.1-1.2\,K and in magnetic fields up to 5\,T.
Both at zero field and at 5\,T the wire exhibits the Wiedemann-Franz law with the fundamental Lorenz number, contrary to an earlier report
[Gloos, K.\ \emph{et al}, Cryogenics \textbf{30}, 14 (1990)].
Our result demonstrates that silver is an excellent material for thermal links in ultra-low-temperature experiments operating at high magnetic fields.}

\maketitle

In metals at low temperature both the electrical and thermal transport is dominated by the conduction electrons undergoing elastic scattering
off impurities and crystallographic defects.
Such processes contribute equally to the momentum and energy transfer and can be described by the elementary concept of mean free path,
leading to the ubiquitous relationship between the electrical and thermal conductivities $\sigma$ and $\kappa$,
the Wiedemann-Franz (WF) law~\cite{Sommerfeld1928,Kumar1993},
\begin{equation}\label{eq:WF}
\kappa = L_0 T \sigma,
\end{equation}
characterised by the fundamental Sommerfeld's value
\begin{equation}\label{eq:L0}
L_0 = \pi^2 k_B^2\big / 3 e^2 = 2.44 \times 10^{-8}\,\mathrm{W\,\Omega\,K^{-2}}
\end{equation}
of the Lorenz number. Here $T$ is the temperature, $k_B$ is the Boltzmann constant and $e$ is the electron charge.
At sufficiently low temperature the WF law is observed in a wide range of conductors~\cite{Macia2009},
including low-dimensional~\cite{Chiatti2006, Jezouin2013} and strongly-correlated~\cite{Machida2013} electron systems.

In addition to superconductors with $\kappa / \sigma = 0$, and exotic Fractional Quantum Hall states with fractional charge,
where $e$ in Eq.~\eqref{eq:L0} is replaced with i.e. $e/3$~\cite{Banerjee2017},
deviations from WF law at low temperatures have been reported in silver and non-superconducting aluminium with residual resistivity ratio $RRR = 400-3000$,
potentially linked to anomalies in the electron scattering in these conventional metals~\cite{Gloos1990}.
The reported experimental Lorenz numbers $L = \kappa / T\sigma$ an order of magnitude below the fundamental value $L_0$ have implications to cryogenic engineering,
where both materials are widely employed: aluminium for superconducting heat switches and silver for high-magnetic-field thermal links.
It is this use of silver, that has motivated our study.

The appeal of silver for high-field ultra-low-temperature applications is the unusually small nuclear magnetic moments
$\mu \sim 0.1\mu_N$, where $\mu_N$ is the nuclear magneton.
Consequently in the regime of small nuclear polarisation, $B/T < k_B / \mu \sim 10^4$\,T/K,
the contribution of the nuclear magnetism to the heat capacity and heat of magnetisation (both $\propto \mu^2$) 
is two orders of magnitude smaller than for most practical metals, such as copper and aluminium where $\mu \approx \mu_N$~\cite{Pobell2007}.
Since silver cools fast and generates little heat when magnetised, it is important to establish
how well this material transfers heat.
This question remains open: beside their observation of the violation of WF law,
Gloos \emph{et al.}~\cite{Gloos1990} review earlier reports of silver both obeying and violating WF law.

\begin{figure}[t!]
\raisebox{0.5em}{\includegraphics[scale=0.85]{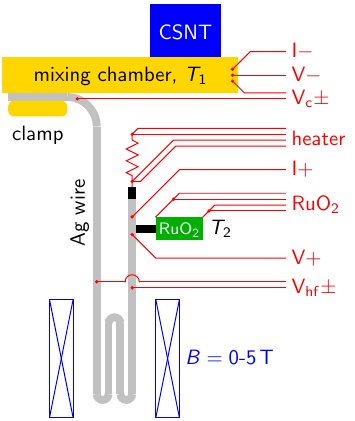}}\hfill%
\parbox[b]{2.5in}{\caption{Experimental setup for combined measurements of electrical and thermal transport in a silver wire.
The wire is clamped to the mixing chamber plate of a dilution refrigerator at a temperature $T_1$, measured
with current-sensing noise thermometer (CSNT).
A constantan heater is attached to the opposite end of the wire.
The temperature $T_2$ of a nearby point is measured with a RuO$_2$ resistance thermometer.
The thermal conductance of the silver wire is inferred from measuring the temperature gradient across the wire as a function of heater power.
The wire passes through a superconducting magnet in order to tune the resistivity of silver with field.
The 4-terminal measurements of the resistance of the entire length of the silver wire sample,
the high-field region and the clamp are conducted using the voltage probes V$\pm$, V$_{\mathrm{hf}}\pm$ and V$_{\mathrm{c}}\pm$,
and shared current leads I$\pm$.}\label{fig:exp}}
\end{figure}

We present the experimental test of WF law in polycrystalline silver with a distribution of $RRR$ between approximately 200 and 400 across the sample.
Our sample is a 1.7\,m long 1\,mm diameter silver wire of 99.99\% purity~\cite{AdventAg}.
As shown in Fig.~\ref{fig:exp}, one end of the wire was clamped to the mixing chamber of a dilution refrigerator
providing good electrical and thermal contact.
In order to measure the thermal conductance, a heater and a RuO$_2$ thermometer were attached to the other end.
The wire was rigidly mounted to a frame made out of sintered alumina and PTFE, with negligible thermal conductance below 1\,K.
From the mixing chamber the wire heads into the bore of a superconducting magnet, where 1\,m of the wire is folded into a meander;
the other end of the wire exits the high-field region,
so the magnetoresistance in the RuO$_2$ thermometer and heater does not need to be taken into account.

The temperature $T_1$ of the mixing chamber was measured by a current-sensing noise thermometer (CSNT)~\cite{Casey2014}.
The CSNT was calibrated against the superconducting transition $T_c = 1.18$\,K in Al/Si1\% wire~\cite{Ventura1998} that forms part of the noise sensor.
The 1\% systematic uncertainty of this calibration propagates to all measurements.
The RuO$_2$ thermometer was calibrated in situ against CSNT between 0.05 and 2\,K,
with zero field in the magnet and the heater disconnected from the power supply.

\begin{figure}
\includegraphics[scale=0.67]{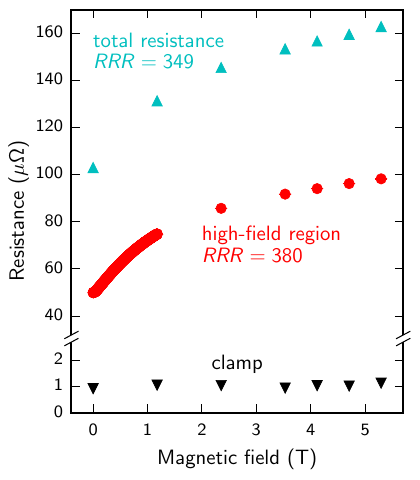}\hfill%
\parbox[b]{2.6in}{\caption{Magnetoresistance in the silver wire below 1\,K.
The resistance of the high-field region grows with field by a factor of 2, leading to a
60\% increase in the total resistance of the wire, that we compare to the measured thermal conductance.
The residual resistivity ratio $RRR$ in zero field is quoted for the entire sample and high-field region alone.
The former is smaller than the latter as a consequence of stronger deformations while installing the experiment. 
The clamp anchoring the wire to the mixing chamber makes a negligible field-independent contribution to the total resistance.}}
\label{fig:RvsB}
\end{figure}

The eight electrical connections to the wire and mixing chamber enable independent 4-terminal measurements of the 
resistances of the entire wire, high-field region and the clamp, illustrated in Fig.~\ref{fig:RvsB}.
The resistances were measured with DC currents up to 9\,mA at room temperature and up to 1\,A below 1\,K.
The RuO$_2$ thermometer and heater are also equipped with 4-terminal connections, allowing precise measurements of their resistances.
To prevent parasitic heat flow, all electrical leads shown in Fig.~\ref{fig:exp} are made out of 0.1\,mm NbTi wires with no resistive matrix,
at least 15\,cm long between the experiment and the heat sink at the mixing chamber plate.

As supplied, the wire has $RRR = 180$. We raised it to $RRR = 440$ by annealing at 800\,$^\circ$C in vacuum for 10 hours,
as measured in a simultaneously annealed witness section of the same wire.
This increase in $RRR$ is easily lost when the wire is bent and crystal defects are reintroduced:
deforming the annealed witness into a spiral with 4\,mm radius of curvature recovered $RRR = 180$.
Due to limited space inside the furnace, the sample wire was folded multiple times during the anneal.
While installing the experiment on the refrigerator, we avoided wire deformation as much as possible and kept the radius of curvature above 4\,mm,
but inevitably the wire was straightened in some places and bent in others.
As a result the sample with overall $RRR = 349$ consists of regions with different $RRR$
in the range $\sim 200$-400 set by the measurements on the witness, presented above.
The high-field section was shaped into the meander prior to annealing, and the installation was less disruptive to it ($RRR = 380$),
than to the rest of the wire ($RRR = 320$).

The wire was pressed against the gold-plated copper mixing chamber plate over a length of 7\,cm.
The clamp resistance of 1\,$\mu\Omega$, Fig.~\ref{fig:RvsB}, includes several mm of the silver wire and may be dominated by this contribution.
Therefore the contact resistance between the sample and the mixing chamber is not measured directly,
but is found to be more than two orders of magnitude smaller than the sample.
This is significant for our 3-terminal thermal conductance measurements.
Shorter clamps were used to attach the copper bobbin hosting the constantan heater wire and the commercial RuO$_2$ package.
The latter is within several mm of the V$+$ contact, therefore the total resistance shown in Fig.~\ref{fig:RvsB} 
can be directly related to the thermal conductance measurements, which we turn to now.

\begin{figure}
\includegraphics[scale=0.67]{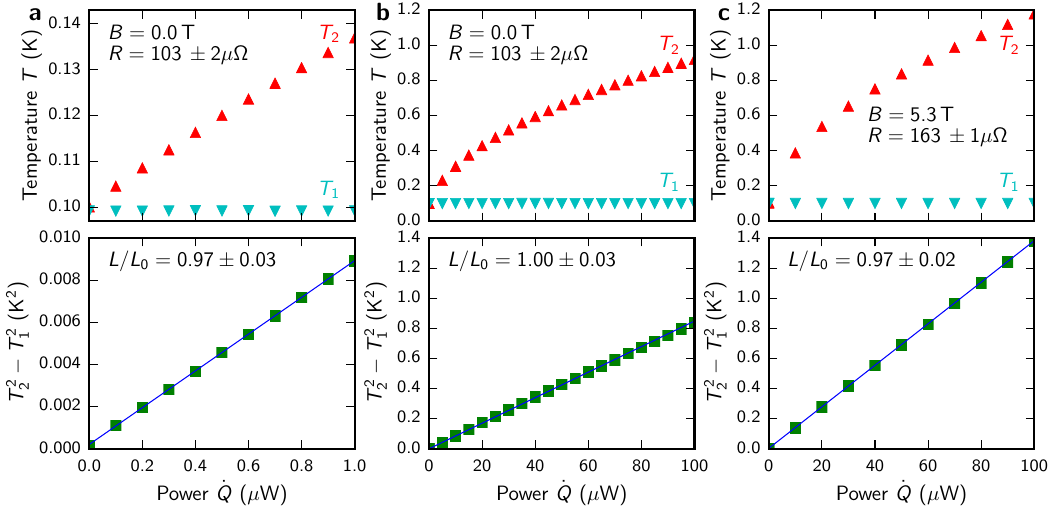}
\caption{Thermal conductance measurements and inferred Lorenz number of the silver wire as a function of magnetic field.
The temperature $T_2$ of the heated end of the wire is measured as a function of heater power at a constant temperature $T_1 = 0.1$\,K of the other end.
Good agreement with Eq.~\eqref{eq:dotQvsT2} is observed near 0.1\,K (\textbf{a}) and up to over 1\,K (\textbf{b}, \textbf{c})
both in zero magnetic field (\textbf{a}, \textbf{b}) and at 5.3\,T (\textbf{c}).
Fits to Eq.~\eqref{eq:dotQvsT2} using the wire resistance $R$ measured in situ, Fig.~\ref{fig:RvsB},
yields the Lorenz number $L$ in excellent agreement with the fundamental value $L_0$, given by Eq.~\eqref{eq:L0}.}\label{fig:L}
\end{figure}

With the mixing chamber stabilised at $T_1 = 0.1$\,K the temperature $T_2$ of the other end of the wire was measured as a function of
power $\dot Q$ dissipated in the heater, shown in Fig.~\ref{fig:L}.
Assuming that the sample thermal conductance follows a power law
\begin{equation}\label{eq:G}
G(T) = A T^\alpha,{\vphantom{\bigg(}}
\end{equation}
the heat flow
\begin{equation}\label{eq:dotQvsTalpha}
\dot Q + \dot Q_0 = \int\limits_{T_1}^{T_2} G(T)\,dT = (\alpha+1)A \left(T_2^{\alpha + 1} - T_1^{\alpha + 1}\right)
\end{equation}
\vskip-0.5em\noindent
has a power-low dependence on $T_1$ and $T_2$.
Here $\dot Q_0$ is the residual heat load on the heater.
Provided the thermal conductivity everywhere in the sample has the same temperature dependence $\kappa \propto T^\alpha$,
the $\dot Q_0$ term describes the heat applied not only directly to the heater, but also to any intermediate point along the sample.
In the latter case the equivalent $\dot Q_0$ is smaller than the actual power dissipated in the experiment.

Our data taken between 0.1 and 1\,K, Fig.~\ref{fig:L}, are consistent with Eq.~\eqref{eq:dotQvsTalpha} and indicate $\alpha + 1 = 2$,
in agreement with Wiedemann-Franz law with temperature-independent electrical resistivity.
We parameterise the prefactor $A = L / R$
via the Lorenz number $L$ and electrical resistance $R$ of the sample and rewrite Eq.~\eqref{eq:dotQvsTalpha} as
\begin{equation}\label{eq:dotQvsT2}
T_2^2 - T_1^2 = \frac{2 R}{L} \left( \dot Q + \dot Q_0\right).{\vphantom{\int\limits_a^b}}
\end{equation}
Fitting this equation to the data and taking $R$ measured in situ, Fig.~\ref{fig:RvsB},
we find that the Lorenz number $L$ lies within 3\% of the fundamental value $L_0$ both in zero field and at 5.3\,T, see Fig.~\ref{fig:L}.

Our in situ calibration of the RuO$_2$ thermometer is susceptible to a systematic error
due to a temperature gradient across the wire caused by parasitic heating.
In the $T_1 = 0.1-1$\,K temperature range of our experiment we estimate this error to be less than 10\,mK, since
the calibration does not saturate down to $T_1 < 50$\,mK.
Moreover, we find that the determination of $A$ and $\alpha$ is immune to an error of this type,
as long as the thermal conductivity has the same temperature dependence $\kappa \propto T^\alpha$ throughout the sample, consistent with our data.

In this experiment WF law is observed down to 0.1\,K.
Since this temperature is firmly in the regime of residual electrical resistivity, we expect no violation of WF law at lower temperatures.
Our sample is comprised of regions with different $RRR$; furthermore, when the magnetic field is applied,
its strength and orientation with respect to the electrical and thermal currents varies across the sample.
Although we only observe the overall fundamental Lorenz number $L = L_0$ in this heterogeneous system,
we argue that this result holds for each individual region,
since the presence of both $L > L_0$ and $L < L_0$ anomalies, that cancel each other out at different values of the magnetic field is highly unlikely.

While our result questions the reported violation of WF law in silver~\cite{Gloos1990}, it is important to recognise
the difference between the experimental conditions, that may lead to the discrepancy.
First, the measurements in Ref.~\cite{Gloos1990} were performed at 2-10\,K, while 
the parasitic thermal conductance through the alumina/PTFE support structure limited our study to $T < 1$\,K.
No significant deviations from the $\alpha = 1$ power law were observed in either experiment,
therefore anomalous temperature dependence $G(T)$ in the narrow intermediate range between 1 and 2\,K,
that would reconcile these results, is unlikely.

Second, magnetic impurities, such as iron, present in high-purity silver at ppm level, may be the dominant scattering centres in our sample.
These impurities are passivated by the oxygen anneal used by Gloos \emph{et al.}~\cite{Gloos1990}, so their work may indicate
an anomalous intrinsic scattering mechanism in silver, that our experiment on a sample annealed in vacuum is insensitive to.
However, doubts are cast on this intriguing scenario by a recent observation of WF law in a zinc heat switch,
where air-annealed silver leads with $RRR = 3000$ constitute roughly half of the total resistance~\cite{Toda2022}.

Future experiments should compare different annealing procedures
and aim for homogeneous measurement conditions, including $RRR$, magnetic field strength and
its orientation with respect to the direction of electrical current and heat flow.
In the absence of such detailed studies we recommend not to anneal in oxygen silver thermal links intended for high-field experiments.
While this process is known to remarkably increase $RRR$~\cite{Ehrlich1974},
the magnetoresistance is strengthened simultaneously, reducing the benefit of high $RRR$ in high fields.

In conclusion, we have demonstrated Wiedemann-Franz law with the fundamental Lorenz number
in high-purity vacuum-annealed polycrystalline silver with $RRR \sim 200$-400 below 1\,K and in magnetic fields up to 5\,T.
Therefore silver, with its practical metallurgical properties and small nuclear magnetic moments,
is an excellent material for thermal links in the regime of high magnetic field $B$
and ultra-low temperature $T$ up to $B/T \sim 10^{4}$\,T/K.

\bmhead{Acknowledgments}
This work was supported by the European Microkelvin Platform, EU’s H2020 project under grant agreement no. 824109.
We thank Richard Elsom, Paul Bamford and Ian Higgs for excellent technical support.


\end{document}